\documentclass[conference]{IEEEtran}
\IEEEoverridecommandlockouts
\usepackage{graphicx}
\usepackage{amssymb,amstext,amsmath}
\usepackage{setspace}
\usepackage{multirow}
\usepackage{makecell}
\usepackage{color}
\usepackage{caption}
\usepackage{verbatim}
\usepackage{float}
\usepackage{mathtools}
\usepackage{amssymb}
\usepackage{algorithm}
\usepackage{color}
\usepackage{booktabs}
\usepackage{tikz}
\usepackage{balance}
\usepackage{mathrsfs}
\usepackage{algorithm,algcompatible,tabularx}
\usepackage{amsthm}
\usepackage{subcaption}
\usepackage{scalerel}
\usepackage{cite}
\usepackage{balance}

\makeatletter
\def\thmheadbrackets#1#2#3{%
  \thmname{#1}\thmnumber{\@ifnotempty{#1}{ }\@upn{#2}}%
  \thmnote{ {\the\thm@notefont[#3]}}}
\makeatother

\newtheoremstyle{mytheoremstyle} 
    {\topsep}                    
    {\topsep}                    
    {\itshape}                   
    {}                           
    {\scshape}                   
    {}                          
    {.5em}                       
    {\thmheadbrackets{#1}{#2}{#3}}  
\theoremstyle{mytheoremstyle}

\newcommand{\x}{{\bf x}}

\newcommand{\y}{{\bf y}}

\newcommand{\z}{{\bf z}}

\newcommand{\bu}{{\bf u}}
\newcommand{\bv}{{\bf v}}

\newcommand{\btheta}{\boldsymbol\theta}
\newcommand{\bbeta}{\boldsymbol\beta}

\newcommand{\bSigma}{\boldsymbol\Sigma}

\newcommand{\bmu}{\boldsymbol\mu}

\makeatletter
\newcommand{\multiline}[1]{%
  \begin{tabularx}{\dimexpr\linewidth-\ALG@thistlm}[t]{@{}X@{}}
    #1
  \end{tabularx}
}

\def\BibTeX{{\rm B\kern-.05em{\sc i\kern-.025em b}\kern-.08em
    T\kern-.1667em\lower.7ex\hbox{E}\kern-.125emX}}

\begin{document}

\title{Fusion of Information in Multiple Particle Filtering in the Presence of Unknown Static Parameters

}

\author{
\IEEEauthorblockN{Xiaokun Zhao, Marija Iloska, Yousef El-Laham, Mónica F. Bugallo}
\IEEEauthorblockA{\textit{Department of Electrical and Computer Engineering} \\
\textit{Stony Brook University}\\
Stony Brook, NY, USA \\
\{xiaokun.zhao, marija.iloska, yousef.ellaham, monica.bugallo\}@stonybrook.edu}
}

\maketitle
\begin{abstract}
An important and often overlooked aspect of particle filtering methods is the estimation of unknown static parameters.
A simple approach for addressing this problem is to augment the unknown static parameters as auxiliary states that are jointly estimated with the time-varying parameters of interest. 
This can be impractical, especially when the system of interest is high-dimensional. 
Multiple particle filtering (MPF) methods were introduced to try to overcome the \emph{curse of dimensionality} by using a “divide and conquer" approach, where the vector of unknowns is partitioned into a set of subvectors, each estimated by a separate particle filter. 
Each particle filter weighs its own particles by using predictions and estimates communicated from the other filters. 
Currently, there is no principled way to implement MPF methods where the particle filters share unknown parameters or states. 
In this work, we propose a fusion strategy to allow for the sharing of unknown static parameters in the MPF setting. 
Specifically, we study the systems which are separable in states and observations. 
It is proved that optimal Bayesian fusion can be obtained for state-space models with non-interacting states and observations.
Simulations are performed to show that MPF with fusion strategy can provide more accurate estimates within fewer time steps comparing to existing algorithms.
\end{abstract}
\section{Introduction}
Particle filtering (PF) is a Monte Carlo methodology that is used for dynamic state estimation in nonlinear and non-Gaussian probabilistic systems \cite{doucet2001introduction,arulampalam2002tutorial}. 
The methodology is based on a classical statistical technique called importance sampling \cite{robert2013monte}, whereby a set of samples (called particles) and weights are used to form a discrete random measure that approximates the posterior distribution of the unknown states given the observations. 
While other filtering methodologies that deal with nonlinear systems have been proposed  (e.g., extended Kalman filtering (EKF) \cite{welch1995introduction} or unscented Kalman filtering (UKF) \cite{wan2000unscented}), 
PF methods can handle arbitrary distributional assumptions about the system of interest. 
This flexibility makes PF powerful for dealing with highly nonlinear and non-Gaussian systems. 

Despite the success of PF in many practical applications \cite{djuric2008target,nakamura2009parameter,gustafsson2002particle}, the performance of the method heavily depends on a number of factors. 
For instance, the presence of unknown static parameters in the probabilistic system of interest introduces some difficulties, since the framework for PF methods is traditionally setup for online state estimation  \cite{liu2001combined}. 
One approach to dealing with this problem is to use Rao-Blackwellization to analytically integrate out the unknown static parameters \cite{li2004estimation}. 
Unfortunately, this can only be done for certain mathematical models. 
In the more general scenario, the typical approach is to augment the static parameter as an unknown state in the probabilistic system. 
One methodology, for instance, is to model the dynamics of the unknown static parameter as a Gaussian random walk with small noise variance \cite{kitagawa1998self}.
Another approach uses a kernel-based approach \cite{liu2001combined}, whereby the posterior distribution of the unknown parameters is approximated with a mixture of Gaussians, whose parameters are updated in each time instant. Another methodology, called \textit{density-assisted particle filtering} (DAPF) \cite{djuric2004density}, approximates the joint posterior of the unknown latent states and static parameters using a parameteric density. Special cases of this approach are Gaussian particle filtering (GPF) \cite{1232326} and Gaussian sum particle filtering (GSPF) \cite{1232327}. 

In addition to the challenge of dealing with static parameters, since PF is a Monte Carlo methodology, it also suffers from the \textit{curse of dimensionality}, which causes the performance of the method to deteriorate as the dimension of the unknown state vector increases \cite{bengtsson2008curse}. 
One way to alleviate the problems posed by the curse of dimensionality is to consider filtering methodologies which partition the state-space of the unknown states.
Multiple particle filtering (MPF) \cite{djuric2007multiple}, for instance, uses the so-called divide and conquer approach to address the curse of dimensionality.  
In MPF, the full state vector is partitioned into substates, each dealt with by a separate particle filter. 
The weighting of the particles is carried out by the exchange of information between filters, which can be done in a variety of ways \cite{djuric2013particle,ait2016variational}. 
It is still unclear how one would also deal with unknown static parameters in the MPF framework, since information about the static parameters may be required across all filters.

In this paper, we propose a novel approach based on Bayesian fusion to tackle the problem of shared unknown static parameters in MPF when the states and observations in the state-space model are non-interacting. 
We provide numerical simulation results to show that the MPF framework with fusion outperforms the standard particle filtering (SPF) and DAPF especially in high dimensional space. 
The estimation also converges faster compared to MPF without information fusion.
\section{Problem Formulation}
\label{sec: problem_formualton}
Let $\x_t\in\mathbb{R}^{d_x}$ denote a $d_x$-dimensional latent state vector, $\y_t\in\mathbb{R}^{d_y}$ a $d_y$-dimensional measurement vector, and $t$ a time-index. 
We consider a state-space model, where the system evolves according to the following set of equations
\begin{align}
    \label{eq: state_space_formulation}
    \x_t&=f(\x_{t-1},\bu_t, \btheta), \\
    \y_t&=g(\x_t,\bv_t, \btheta),
\end{align}
where $f(\cdot)$ is the system transition function, $\bu_t$ is the state noise that has a distribution $p(\bu_t)$, $g(\cdot)$ is the measurement function, $\bv_t$ is the measurement noise that has a distribution $p(\bv_t)$, and $\btheta\in\mathbb{R}^{d_\theta}$ is a $d_\theta$-dimensional unknown static parameter vector that parameterizes both $f(\cdot)$ and $g(\cdot)$. 

For a given time $t$, our goal is to jointly estimate the latent state $\x_t$ and the static parameter vector $\btheta$ given the set of measurements $\y_{1:t}\triangleq\{\y_1,\y_2,\ldots,\y_t\}$ under the Bayesian paradigm. This amounts to determining the joint posterior distribution of $\x_t$ and $\btheta$
\begin{align}
    p(\x_t,\btheta|\y_{1:t})=p(\x_t|\btheta,\y_{1:t})p(\btheta|\y_{1:t}).
\end{align}
The distribution $p(\x_t|\btheta,\y_{1:t})$ represents the filtering distribution under the assumption that the static parameter vector is known, while $p(\btheta|\y_{1:t})$ is the posterior distribution of $\btheta$.
\section{Multiple Particle Filter with Shared Unknown Parameters}
\label{sec: general_strategy}
In this section, we propose a novel MPF algorithm that fuses information across filters in cases that filters share parameters for systems where the states and observations are non-interacting.

Let the state vector $\x_t$ be partitioned as $\x_t=[\x_{1,t}^\intercal,\x_{2,t}^\intercal,\ldots,\x_{K,t}^\intercal]^\intercal$ and let the unknown static parameter vector $\btheta$ be partitioned as $\btheta=[\btheta_{1,\ell}^\intercal,\btheta_{2,\ell}^\intercal,\ldots, \btheta_{K,\ell}^\intercal,\btheta_{g}^\intercal]^\intercal$.
Since the states and observations are non-interacting, after partitioning, the $k$th subsystem of the state-space model can be represented via the following set of equations:
\begin{align}
    \label{eq: state_equation_study_1}
    \x_{k,t}&=f_k(\x_{k,t-1},\bu_{k,t}, \btheta_{k,\ell}, \btheta_{g}),\quad  k=1,\ldots, K\\
    \label{eq: obs_equation_study_1}
    \y_{k,t}&=g_k(\x_{k,t},\bv_{k,t}, \btheta_{k,\ell}, \btheta_{g}),\quad  k=1,\ldots, K,
\end{align}
where $f_k(\cdot)$ and $g_k(\cdot)$ are the system transition function and measurement function local to the $k$th subsystem, $\btheta_{k,\ell}$ stands for the local parameter vector in the $k$th subsystem, and $\btheta_{g}$ is the global parameter vector. 
In this type of system, the states and observations contained in each subsystem are independent of the states and observations from all other subsystems when the global parameter $\btheta_g$ is known. 

In MPF, $K$ particle filters are used to jointly estimate the unknowns $\x_t$ and $\btheta$, where each filter tracks one of the subsystems above. 
In this setting, we consider that the $k$th filter jointly tracks $\z_{k,t}\triangleq\{\x_{k,t},\btheta_{k,\ell},\btheta_{g}\}$. 
At each time instant $t$, the $k$th particle filter represents an approximation of the joint posterior distribution $p(\z_{k,t}|\y_{1:t})$ using a set of particles and weights
\begin{equation}
    \label{eq: posterior_approx_jth_unfused}
    \hat p(\z_{k,t}|\y_{1:t}) = \sum_{n=1}^{N_k} w_{k,t}^{(n)}\delta(\z_{k,t}^{(n)}-\z_{k,t}),
\end{equation}
where $\z_{k,t}^{(n)}\triangleq\{\x_{k,t}^{(n)},\btheta_{k,\ell}^{(n)},\btheta_{g}^{(n,k)}\}$ is the $n$th particle of the $k$th filter. 
Using the chain rule of probability theory, we can write the posterior distribution $p(\z_{k,t}|\y_{1:t})$ as
\begin{equation}
    \label{eq: posterior_approx_jth_cond}
    p(\z_{k,t}|\y_{1:t})=p(\x_{k,t},\btheta_{k,\ell}|\btheta_g, \y_{1:t})p(\btheta_g|\y_{1:t}).  
\end{equation}
This essentially means that each filter has an approximation of the marginal posterior of the global parameters $p(\btheta_g|\y_{1:t})$. We denote the approximation of $p(\btheta|\y_{1:t})$ according to the $k$th filter as $\hat p_k(\btheta|\y_{1:t})$ and it is given by
\begin{equation}
    \label{eq: p_theta_g_approx}
    \hat p_k(\btheta_g|\y_{1:t}) = \sum_{n=1}^{N_k} w^{(n)}_{k,t} \delta(\btheta_g-\btheta_g^{(n,k)}).
\end{equation}
Since each filter has its own approximation to $p(\btheta_g|\y_{1:t})$, we would like to fuse these approximations to obtain a better/more accurate distribution $q(\btheta_g|\y_{1:t})$. 
Given the fused distribution, we can then update the $k$th filter's full posterior accordingly by replacing $p(\btheta_g|\y_{1:t})$  in \eqref{eq: posterior_approx_jth_cond} with  $q(\btheta_g|\y_{1:t})$:
\begin{equation}
    \label{eq: posterior_approx_jth_cond_fuse}
    q(\z_{k,t}|\y_{1:t})=p(\x_{k,t},\btheta_{k,\ell}|\btheta_g, \y_{1:t})q(\btheta_g|\y_{1:t}).
\end{equation}
The update rule in \eqref{eq: posterior_approx_jth_cond_fuse} can be implemented as a resampling step in the MPF framework. This is shown in Algorithm \ref{alg: MPF_basic_fusion_strategy}.

\begin{algorithm}[t]
\caption{Fusion Strategy for Multiple Particle Filtering}
\label{alg: MPF_basic_fusion_strategy}
  {\bf Input}: $\{\z_{1,t}^{(n)},w_{1,t}^{(n)}\}_{n=1}^{N_1},\ldots,\{\z_{K,t}^{(n)},w_{K,t}^{(n)}\}_{n=1}^{N_K}$\\
  {\bf Output}: $\{\tilde\z_{1,t}^{(n)}\}_{n=1}^{N_1},\ldots,\{\tilde\z_{K,t}^{(n)}\}_{n=1}^{N_K}$ 
  \begin{algorithmic}[1]
    \FOR{$k=1,\ldots,K$}
        \STATE \multiline{Approximate $p(\z_{k,t}|\y_{1:t})$ with a parametric density $\tilde p(\z_{k,t} ; \bbeta_{k,t})$ by estimating the parameters $\bbeta_{k,t}$ using the set of samples and weights $\{\z_{k,t}^{(n)},w_{k,t}^{(n)}\}_{n=1}^{N_k}$. 
        }
    \ENDFOR
    \STATE Obtain $ q(\btheta_g|\y_{1:t})$ by fusing the marginal distributions by \eqref{eq: fusion_function_update}
    \FOR{$k=1,\ldots,K$}
        \STATE Sample $N_k$ particles from the fused marginal distribution by \eqref{eq: draw_particles_fused}
        \FOR{$n=1,\ldots,N_k$}
        \STATE Sample a particle from the conditional distribution by \eqref{eq:resample_local}
        \STATE Set $\tilde \z_{k,t}^{(n)} =  \left\{\tilde \x_{k,t}^{(n)},\tilde\btheta_{k,\ell}^{(n)}, \tilde \btheta_{g}^{(n,k)}\right\}$. 
        \ENDFOR
    \ENDFOR
  \end{algorithmic}
\end{algorithm}

The idea of Algorithm \ref{alg: MPF_basic_fusion_strategy} is to use the set of samples and weights from each of the particle filters and to return to each of those particle filters a set of resampled particles. The resampling of the particles is based on fusing the information from the filters. In order to accomplish this fusion in a principled way, we approximate the current filtering density of each particle filter using a parametric density
\begin{equation}
    \label{eq: approximation_parametric}
    \tilde p(\z_{k,t}; \bbeta_{k,t}) \approx p(\z_{k,t}|\y_{1:t}),
\end{equation}
where the parameters $\bbeta_{k,t}$ are estimated using set of samples and weights from the $k$th particle filter, $\{\z_{k,t}^{(n)}, w_{k,t}^{(n)}\}_{n=1}^{N_k}$. The idea is to choose the parametric density $\tilde p(\z_{k,t}; \bbeta_{k,t})$ from a family of probability distributions such that: 
\begin{enumerate}
    \item The marginal density $\tilde p(\btheta_g|\bbeta_{k,t})$ can be obtained analytically.
    \item The conditional density $\tilde p(\x_{k,t},\btheta_{k,\ell}|\tilde \btheta_{g}; \bbeta_{k,t})$ can be obtained analytically.
\end{enumerate}
One parametric family of distributions that is known to satisfy the above requirements is the Gaussian distribution. In this case $\bbeta_{k,t}\triangleq\{\bmu_{k,t}, \bSigma_{k,t}\}$, where $\bmu_{k,t}$ and $\bSigma_{k,t}$ correspond to the mean and covariance matrix of that Gaussian. 

After, estimating the parameters $\bbeta_{k,t}$, we marginalize the approximate filtering densities in order to obtain $K$ different  approximations to the posterior distribution of the global parameters $\btheta_g$. 
The goal is to fuse these $K$ approximate posteriors into a single probability distribution.
This fused distribution $q(\btheta_g|\y_{1:t})$ is obtained by using a fusion function $h : \mathcal{P}^K\rightarrow\mathcal{P}$\cite{koliander2022fusion}:
\begin{equation}
    \label{eq: fusion_function}
    q(\btheta_g|\y_{1:t})=h\left(\tilde p(\btheta_g; \bbeta_{1,t}),\ldots, \tilde p(\btheta_g; \bbeta_{K,t})\right),
\end{equation}
where $\mathcal{P}$ denotes the set of all valid probability density functions for the random vector $\btheta_g$.

In terms of choosing the fusion function for the system represented by \eqref{eq: state_equation_study_1}-\eqref{eq: obs_equation_study_1}, the fusion function which allows for optimal Bayesian fusion can be obtained. 
We refer to optimal Bayesian fusion as a fusion rule that can fuse the local posterior distributions of the global parameter vectors (given their latest local measurement) into the posterior distribution of the global parameter vector given \emph{all measurements} up to current time instant.
To get the approximation of global posterior distribution $p(\btheta_g|\y_{1:t})$, we combine all the approximations of the local posterior distributions $p(\btheta_g|\y_{k,t}, \y_{1:t-1})$ learned from each filter and the global posterior distribution  $p(\btheta_g|\y_{1:t-1})$ learned from last time step.

At a particular time instant $t$, we can write the marginal posterior distribution of $\btheta_g$ given $\y_{1:t}$ as 
\begin{equation}
    p(\btheta_g|\y_{1:t})\propto p(\y_t|\btheta_g,\y_{1:t-1})p(\btheta_g|\y_{1:t-1}).
\end{equation}
In a state-space model described by \eqref{eq: state_equation_study_1}-\eqref{eq: obs_equation_study_1}, the likelihood $p(\y_t|\btheta_g,\y_{1:t-1})$ can be proven to be the product of likelihoods of all subsystems
\begin{align}
    p(\y_t|\btheta_g,\y_{1:t-1})
    &=\prod_{k=1}^K p(\y_{k,t}|\btheta_g, \y_{1:t-1}).
\end{align}
Applying Bayes' theorem, we can rewrite the likelihood of $k$th filter as
\begin{align}
    p(\y_{k,t}|\btheta_g,\y_{1:t-1})
    \propto\frac{p(\btheta_g|\y_{k,t}, \y_{1:t-1})}{p(\btheta_g|\y_{1:t-1})}.
\end{align}
Thus, we can get the optimal Bayesian fusion
\begin{align}
    \label{fuse}
    p(\btheta_g|\y_{1:t})\propto 
    \frac{\prod_{k=1}^K p(\btheta_g|\y_{k,t}, \y_{1:t-1})}{p(\btheta_g|\y_{1:t-1})^{K-1}}.
\end{align}
The result in \eqref{fuse} is rather intuitive and has a counterpart for static (time-invariant) systems. 
Here, the posterior distribution $p(\btheta_g|\y_{1:t-1})$ from last time instant serves as a prior distribution for $\btheta_g$ at time instant $t$. 
After each filter processes its local measurements to obtain an updated posterior $p(\btheta_g|\y_{k,t}, \y_{1:t-1})$ based on the prior $p(\btheta_g|\y_{1:t-1})$. 
A correction needs to be made since each filter utilized the prior and a simple multiplication of the $K$ posteriors would count the prior multiple times. This correction is made by multiplying the $K$ posteriors and dividing out the prior $K-1$ times. Then the marginal posterior $q(\btheta_g|\y_{1:t})$ in \eqref{eq: fusion_function} can be obtained:
\begin{equation}
    \label{eq: fusion_function_update}
    q(\btheta_g|\y_{1:t})=\frac{\prod_{k=1}^K \tilde p(\btheta_g; \bbeta_{k,t})}{q(\btheta_g|\y_{1:t-1})^{K-1}}.
\end{equation}
When we assume the distributions are all Gaussian distributions, we can get the mean and covariance matrix of the fused distribution $q(\btheta_g|\y_{1:t})=\mathcal{N}(\bmu_{q,t}, \bSigma_{q,t})$:
\begin{align}
    \bmu_{q,t}&=\bSigma_{q,t}\Big (\sum_{k=1}^K \bSigma_{k,t}^{-1}\bmu_{k,t}-(K-1)\bSigma_{q,t-1}^{-1}\bmu_{q,t-1}\Big ) , \\
    \bSigma_{q,t}&=\Big (\sum_{k=1}^K \bSigma_{k,t}^{-1}-(K-1)\bSigma_{q,t-1}^{-1}\Big 
 )^{-1}.
\end{align} 
In obtaining the fused marginal posterior $q(\btheta_g|\y_{1:t})$, we can now communicate this density to each of the particle filters so that they may use it in their resampling step. For each filter, a set of particles is drawn from the fused marginal posterior
\begin{equation}
    \label{eq: draw_particles_fused}
    \tilde \btheta_{g}^{(n,k)}\sim  q(\btheta_g|\y_{1:t}), \quad n=1,\ldots,N_k.
\end{equation}
We then use the conditional distribution $\tilde p(\x_{k,t},\btheta_{k,\ell}|\tilde \btheta_{g}; \bbeta_{k,t})$ to sample corresponding substates and local parameters:
\begin{equation}
\label{eq:resample_local}
    \left\{\tilde \x_{k,t}^{(n)},\tilde\btheta_{k,\ell}^{(n)}\right\}\sim \tilde p(\x_{k,t},\btheta_{k,\ell}|\tilde \btheta_{g}^{(n,k)}; \bbeta_{k,t}), \quad n=1,\ldots,N_k.
\end{equation}
Using this strategy, each particle filter obtains a set of resampled particles $\{\tilde \z_{k,t}^{(n)}\}_{n=1}^{N_k}$, where $\tilde \z_{k,t}^{(n)} \triangleq \{\tilde \x_{k,t}^{(n)},\tilde\btheta_{k,\ell}^{(n)}, \tilde \btheta_{g}^{(n,k)}\}$. 

\section{Simulation Results}
In this section, we present numerical results with different scaled systems. 
Four different algorithms are tested: SPF with a Gaussian random walk model \cite{kitagawa1998self}; DAPF \cite{djuric2004density}; MPF without fusion, where each partition performs as a SPF and the global estimate is given by the mean of estimates from each partition; MPF with fusion, which is the proposed method in this work. 

For simplicity, we assume that there are not local unknown parameters. Specifically, we consider a system which can be described by the following transition function and measurement function:
\begin{align}
    \label{eq:simulation_systemx}
    x_{i,t} &= \frac{\theta_1}{1+e^{-x_{i, t-1}+\theta_5}}+\theta_2+u_{i,t}, &\quad  i=1,\ldots, d_{\bf x}\\
    \label{eq:simulation_systemy}
    y_{i,t} &= \theta_3x_{i,t}+\theta_4+v_{i,t}, &\quad  i=1,\ldots, d_{\bf x}
\end{align}
where $d_{\bf x}=10$, $u_{i,t}\sim\mathcal{N}(0,\sigma^2_{u_i})$ and $v_{i,t}\sim\mathcal{N}(0,\sigma^2_{v_i})$ are independent and identically distributed zero-mean Gaussian noise with variance $\sigma^2_{u_i}=2$ and $\sigma^2_{v_i}=1$, respectively.
$\btheta_g=[\theta_1, \theta_2, \theta_3, \theta_4, \theta_5]^\intercal=[2, -2, 2, -2, 3]^\intercal$ is a global parameter vector. 
We test all algorithms with different $d_{\btheta_g}$, where the first $d_{\btheta_g}$ parameters are assumed to be unknown parameters to estimate, while the rest are known. 
For further clarification, when $d_{\btheta_g}=j$, we assume that only $\theta_1, \ldots, \theta_j$ are unknown parameters that will be estimated and $\theta_{j+1}, \ldots, \theta_5$ are known. 

The number of particles used in the simulation is $100d_{\btheta_g}$ per state dimension. 
In other words, when $d_{\btheta_g}=2$, we estimate 2 unknown parameters. The number of particles in SPF and DAPF are both $N=2000$. 
In MPF, with the number of partitions being $K=5$, each particle filter utilizes $N_k=400$ particles. 
The initial particles are drawn as follows: $\x^{(n)}_0\sim\mathcal{N}(\boldsymbol 0,\sigma^2_{u_i}\boldsymbol{I})$, $\btheta_g\sim\mathcal{N}(\boldsymbol 1,2\boldsymbol{I})$. 

\begin{figure}
     \centering
     \begin{subfigure}[b]{0.48\textwidth}
         \centering
         \includegraphics[width=1\textwidth]{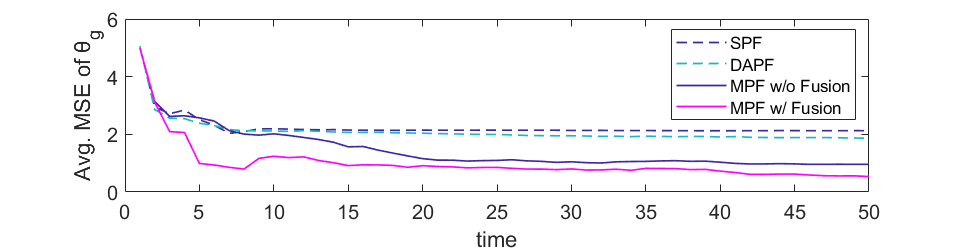}
     \end{subfigure}
     \hfill
     \begin{subfigure}[b]{0.48\textwidth}
         \centering
         \includegraphics[width=1\textwidth]{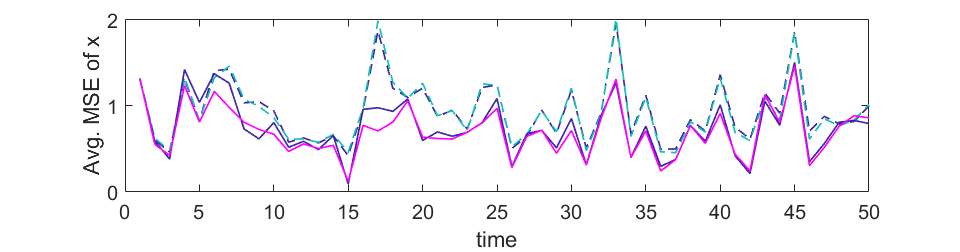}
     \end{subfigure}
        \caption{Averaged MSE of parameter and state estimates with $d_{\bf x}=10$ and $d_{\btheta_g}=2$.}
        \label{fig_d2}
\end{figure}

Figure \ref{fig_d2} shows the average mean square errors (MSEs) of the unknown static parameters and the latent states calculated from 100 realizations of each algorithm with $d_{\btheta_g}=2$. 
In each realization, the system iterates $T=50$ times. 
It is obvious that MPF methods outperform the single particle filter methods (SPF and DAPF) and give lower MSEs on both state and parameter estimation. 
Though the two MPF methods seem to land to a comparable result after 50 iterations, the MPF with information fusion  coverges a lot faster than the MPF without fusion.

\begin{figure}
     \centering
     \begin{subfigure}[b]{0.24\textwidth}
         \centering
         \includegraphics[width=\textwidth]{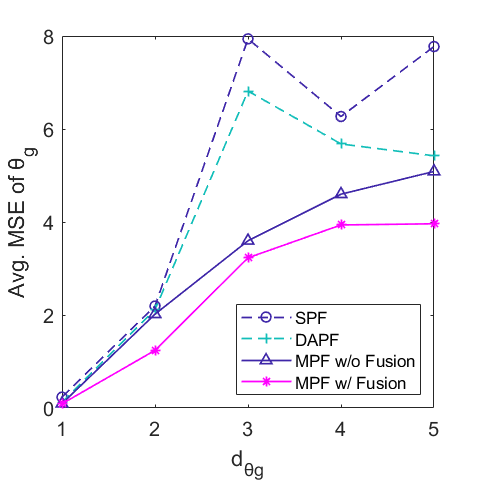}
         \subcaption{$t=5$}
         \label{fig_tmse_t=5}
     \end{subfigure}
     \hfill
     \begin{subfigure}[b]{0.24\textwidth}
         \centering
         \includegraphics[width=\textwidth]
         {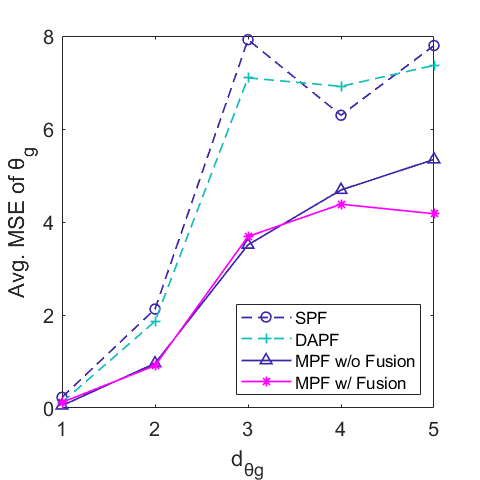}
         \subcaption{$t=50$}
     \end{subfigure}
        \caption{Averaged MSE of the static parameter estimates under different $d_{\btheta_g}$ and $t$ with $d_{\bf x}=10$.}
        \label{fig_tmse_dt}
\end{figure}

The effect of $d_{\btheta_g}$ can be found in Figure \ref{fig_tmse_dt}. Since $\btheta_g$ cannot be partitioned in any of these methods, when the dimensions of unknown static parameters grow larger, all methods will perform worse and provide higher MSEs. However, with the fusion strategy, the performance of our work degrades less than other methods. Specifically, our work can provide a satisfying estimate within only a few time steps as shown in Figure \ref{fig_tmse_t=5}.

\begin{figure}
     \centering
     \begin{subfigure}[b]{0.48\textwidth}
         \centering
         \includegraphics[width=1\textwidth]{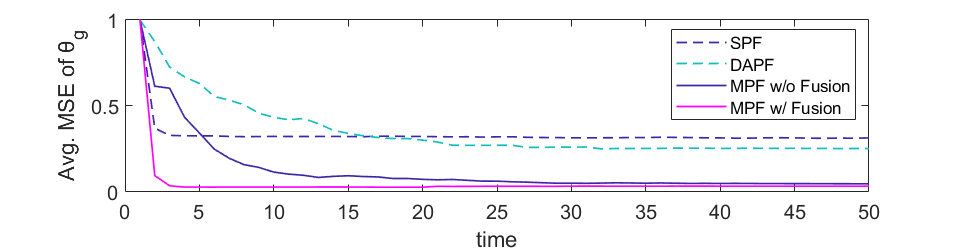}
         \subcaption{}
     \end{subfigure}
     \hfill
     \centering
     \begin{subfigure}[b]{0.24\textwidth}
         \centering
         \includegraphics[width=\textwidth]{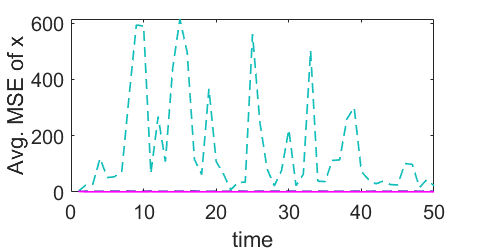}
         \subcaption{}
         \label{mselarge}
     \end{subfigure}
     \hfill
     \begin{subfigure}[b]{0.24\textwidth}
         \centering
         \includegraphics[width=\textwidth]
         {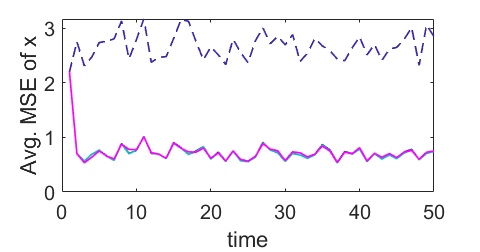}
         \subcaption{}
         \label{msesmall}
     \end{subfigure}
        \caption{Averaged MSE of parameter and state estimates with $d_{\bf x}=100$ and $d_{\btheta_g}=1$. Figure (c) shows the invisible MSEs in Figure (b) due to the scale.}
        \label{fig_d1}
\end{figure}
The system tested above is a relatively smaller system with $d_{\bf x}=10$, which PFs can still handle. However, due to the \textit{curse of dimensionality}, PFs cannot even provide reasonable estimates when the scale of the system becomes large. 
We provide the following results for the same system with $d_{\bf x}=100$ and $d_{\btheta_g}=1$. The number of particles per dimension remains the same.
With more state dimensions, we choose $K=50$. 
Still, 100 realizations are run and the MSEs are recorded in Figure \ref{fig_d1}. 
In such large system, the SPF and DAPF are not able to estimate the latent states properly. 
\section{conclusion}
In this paper we propose a novel fusion strategy used in MPF framework. 
It allows for the sharing of unknown static parameters for state-space models that can be partitioned into independent subsystems. 
We prove that optimal Bayesian fusion can be obtained in this case and the information fusion can be implemented as a resampling step in MPF.
Simulations are run to show that the algorithm give better estimation for both latent states and unknown parameters comparing to the SPF, DAPF, and MPF without fusion strategy.
Since the fusion function is only given for independent subsystems, it remains as future work to derive the approximation of best fusion results in models where the states and observations are interacting.
\balance

\bibliographystyle{IEEEtran}
\bibliography{IEEEabrv, refs}

\begin{thebibliography}{10}
\providecommand{\url}[1]{#1}
\csname url@samestyle\endcsname
\providecommand{\newblock}{\relax}
\providecommand{\bibinfo}[2]{#2}
\providecommand{\BIBentrySTDinterwordspacing}{\spaceskip=0pt\relax}
\providecommand{\BIBentryALTinterwordstretchfactor}{4}
\providecommand{\BIBentryALTinterwordspacing}{\spaceskip=\fontdimen2\font plus
\BIBentryALTinterwordstretchfactor\fontdimen3\font minus \fontdimen4\font\relax}
\providecommand{\BIBforeignlanguage}[2]{{%
\expandafter\ifx\csname l@#1\endcsname\relax
\typeout{** WARNING: IEEEtran.bst: No hyphenation pattern has been}%
\typeout{** loaded for the language `#1'. Using the pattern for}%
\typeout{** the default language instead.}%
\else
\language=\csname l@#1\endcsname
\fi
#2}}
\providecommand{\BIBdecl}{\relax}
\BIBdecl

\bibitem{doucet2001introduction}
A.~Doucet, N.~De~Freitas, and N.~Gordon, ``An introduction to sequential {Monte Carlo} methods,'' in \emph{Sequential Monte Carlo Methods in Practice}.\hskip 1em plus 0.5em minus 0.4em\relax Springer, 2001, pp. 3--14.

\bibitem{robert2013monte}
C.~Robert and G.~Casella, \emph{Monte Carlo Statistical Methods}.\hskip 1em plus 0.5em minus 0.4em\relax Springer Science \& Business Media, 2013.

\bibitem{welch1995introduction}
G.~Welch, ``An introduction to the {K}alman filter,'' 1995.

\bibitem{wan2000unscented}
E.~A. Wan and R.~Van Der~Merwe, ``The unscented {K}alman filter for nonlinear estimation,'' in \emph{Proceedings of the IEEE 2000 Adaptive Systems for Signal Processing, Communications, and Control Symposium (Cat. No. 00EX373)}.\hskip 1em plus 0.5em minus 0.4em\relax Ieee, 2000, pp. 153--158.

\bibitem{arulampalam2002tutorial}
M.~S. Arulampalam, S.~Maskell, N.~Gordon, and T.~Clapp, ``A tutorial on particle filters for online nonlinear/non-{G}aussian {B}ayesian tracking,'' \emph{IEEE Transactions on Signal Processing}, vol.~50, no.~2, pp. 174--188, 2002.

\bibitem{djuric2008target}
P.~M. Djuric, M.~Vemula, and M.~F. Bugallo, ``Target tracking by particle filtering in binary sensor networks,'' \emph{IEEE Transactions on Signal Processing}, vol.~56, no.~6, pp. 2229--2238, 2008.

\bibitem{gustafsson2002particle}
F.~Gustafsson, F.~Gunnarsson, N.~Bergman, U.~Forssell, J.~Jansson, R.~Karlsson, and P.~Nordlund, ``Particle filters for positioning, navigation, and tracking,'' \emph{IEEE Transactions on Signal Processing}, vol.~50, no.~2, pp. 425--437, 2002.

\bibitem{nakamura2009parameter}
K.~Nakamura, R.~Yoshida, M.~Nagasaki, S.~Miyano, and T.~Higuchi, ``Parameter estimation of in silico biological pathways with particle filtering towards a petascale computing,'' in \emph{Biocomputing 2009}.\hskip 1em plus 0.5em minus 0.4em\relax World Scientific, 2009, pp. 227--238.

\bibitem{liu2001combined}
J.~Liu and M.~West, ``Combined parameter and state estimation in simulation-based filtering,'' in \emph{Sequential Monte Carlo Methods in Practice}.\hskip 1em plus 0.5em minus 0.4em\relax Springer, 2001, pp. 197--223.

\bibitem{djuric2007multiple}
P.~M. Djuric, T.~Lu, and M.~F. Bugallo, ``Multiple particle filtering,'' in \emph{2007 IEEE International Conference on Acoustics, Speech and Signal Processing-ICASSP'07}, vol.~3.\hskip 1em plus 0.5em minus 0.4em\relax IEEE, 2007, pp. III--1181.

\bibitem{djuric2004density}
P.~M. Djuric, M.~F. Bugallo, and J.~M{\'\i}guez, ``Density assisted particle filters for state and parameter estimation,'' in \emph{2004 IEEE International Conference on Acoustics, Speech, and Signal Processing}, vol.~2.\hskip 1em plus 0.5em minus 0.4em\relax IEEE, 2004, pp. ii--701.

\bibitem{li2004estimation}
P.~Li, R.~Goodall, and V.~Kadirkamanathan, ``Estimation of parameters in a linear state space model using a {Rao-Blackwellised} particle filter,'' \emph{IEE Proceedings-Control Theory and Applications}, vol. 151, no.~6, pp. 727--738, 2004.

\bibitem{bengtsson2008curse}
T.~Bengtsson, P.~Bickel, B.~Li \emph{et~al.}, ``Curse-of-dimensionality revisited: Collapse of the particle filter in very large scale systems,'' in \emph{Probability and Statistics: Essays in Honor of David A. Freedman}.\hskip 1em plus 0.5em minus 0.4em\relax Institute of Mathematical Statistics, 2008, pp. 316--334.

\bibitem{djuric2013particle}
P.~M. Djuri{\'c} and M.~F. Bugallo, ``Particle filtering for high-dimensional systems,'' in \emph{2013 5th IEEE International Workshop on Computational Advances in Multi-Sensor Adaptive Processing (CAMSAP)}.\hskip 1em plus 0.5em minus 0.4em\relax IEEE, 2013, pp. 352--355.

\bibitem{ait2016variational}
B.~Ait-El-Fquih and I.~Hoteit, ``A variational {B}ayesian multiple particle filtering scheme for large-dimensional systems,'' \emph{IEEE Transactions on Signal Processing}, vol.~64, no.~20, pp. 5409--5422, 2016.

\bibitem{kitagawa1998self}
G.~Kitagawa, ``A self-organizing state-space model,'' \emph{Journal of the American Statistical Association}, pp. 1203--1215, 1998.

\bibitem{1232326}
J.~Kotecha and P.~Djuric, ``Gaussian particle filtering,'' \emph{IEEE Transactions on Signal Processing}, vol.~51, no.~10, pp. 2592--2601, 2003.

\bibitem{1232327}
------, ``Gaussian sum particle filtering,'' \emph{IEEE Transactions on Signal Processing}, vol.~51, no.~10, pp. 2602--2612, 2003.

\bibitem{koliander2022fusion}
G.~Koliander, Y.~El-Laham, P.~M. Djuri{\'c}, and F.~Hlawatsch, ``Fusion of probability density functions,'' \emph{Proceedings of the IEEE}, vol. 110, no.~4, pp. 404--453, 2022.

\end{thebibliography}

\end{document}